\documentclass{JHEP3}

\usepackage{backref}

\title{Peer review and journal models}

\author{Paolo Dall'Aglio\\
Sissa Medialab s.r.l.\\
E-mail: \email{aglio@medialab.sissa.it}}

\abstract{Tentative analysis of alternative peer review and journal
models.  Open Access and Impact factor issues are not
covered. The bibliography, in alphabetical order, lists articles
and web sites (with brief description).}

\begin{document}

\section{Introduction}

The debate about concerning changes in scholarly publishing began well
before the advent of the Internet, but it was with widespread use of
the World Wide Web that many began to consider that a big change was
behind the corner and that in a few year nothing was going to be as
before~\cite{odlyzko}.

In the late nineties several experiments of online journals emerged to
propose new publishing models to make the most of the new medium. A
big emphasis was on the chance to reduce costs and face the so-called
`serials crisis''.

The two main sections analyze journal forms
(section~\ref{journalforms}) and peer review
(section~\ref{peerreview}). The most important content is in the
bibliography that contains articles and web sites in alphabetical
order with back references to the text. All web sites have been
visited in the first half of august 2006, and almost all web sites
references are briefly described.

\subsection{Open Access}

Open Access is not a goal of this study, but it has something in
common with al the subjects covered here.

It is by far the most frequently discussed issue in specialized and
general journals, with several congresses and journal special issues,
not to mention single articles and contributions.

\pagebreak[3]

Just to give an idea we sould mention 
\begin{itemize}
\item The \emph{Open Access Special} on  the journal \emph{Research Information} June/July 2006\\
{\small\href{http://www.researchinformation.info/features_junjul06.html}{\tt http://www.researchinformation.info/features\_junjul06.html}}.

\item SPARC Open Access Newsletter: {\small\href{http://www.earlham.edu/~peters/fos/}{\tt http://www.earlham.edu/~peters/fos/}}.

\item Nature debate on Open Access\\
{\small\href{http://www.nature.com/nature/focus/accessdebate/index.html}{\tt http://www.nature.com/nature/focus/accessdebate/index.html}}.

\item Open Acces Webliography {\small\href{http://dlist.sir.arizona.edu/1005/01/oaw.htm}{\tt http://dlist.sir.arizona.edu/1005/01/oaw.htm}}.

\item Open Access Bibliography:
liberating scholarly literature with e-prints and open access journals
{\small\href{http://www.digital-scholarship.com/oab/oab.htm}
{\tt http://www.digital-scholarship.com/oab/oab.htm}}.

\end{itemize}

Hence, Open Access issues and more generally the problem of costs will
not be covered and will be mentioned only when worth doing.

\section{The ongoing dabate}

The debate is very lilvely and concerns all aspects of scholarly
publishing: peer review, open access, copyright issues, self
archiving. It features all kinds of contributions: from very abstract
to the description of existing services.

This debate is going on mainly on a number of dedicated journals like:
Learned Publishing~\cite{learned},
D-Lib Magazine~\cite{dlib},
Science \& Technology Libraries~\cite{s&tl},
and the no longer maintained Journal of Electronic Publishing~\cite{jep}.

Besides these publications, learned societies and big associations are
funding research on the subject. For example the Association of
Research Libraries~\cite{arl} and the Scholarly Communications Group
of JISC~\cite{scg}.

Also many specialized journals are greatly involved. First of all
Nature, where articles on the subject appear very often and a debate on
peer review is going on \cite{nature-debate-pr} together with an
experiment of open-commentary peer review.

Then journals like BioMed Central~\cite{BMC}, JAMA~\cite{jama} and
BMJ~\cite{BMJ} often publish contributions on the subject, with a
particular attention to the evaluation of their peer review systems.

It must be noted that great part of the discussion involves the
biomedical sector, both scientists and journals staff, and this is
probably due to the high importance those sectors give to quality
controls and results certification.

An invaluable resource is the Scholarly Electronic Publishing
Bibliography~\cite{sepb} by Charles~W.~Bailey~Jr., updated at least
twice a month, lists article, books and online resources.

\section{The journal form}\label{journalforms}

Unlike peer review, that we will discuss later, the debate on the
journal form did not produce a great amount of statistical and
objective studies, mainly because it is not possible give measurable
answers to the basic questions. Do journals work as they are? Do they
need to be modified?

We could divide the contibutions in this field into two categories:
theoretical and experimental.

\subsection{Abstract proposals}

In this first class we can consider articles and studies that devise a
revolutionary approach, without trying it in practice. Such entirely
new models of scholarly communication usually include also a new form
of review.

Few authors have tried to examine abstractly what are (and what should
be) the functions played by a journal and which roles are in charge of
them.

A very abstract study is the one by B.-C.~Bj\" ork~\cite{bjork} that
tries to give a formal model of the scientific communication process.
The model is very detailed and hierarchical and includes the whole
communication chain, from initial research to the assimilation of
research results to everyday practice. Although the model treats both
informal and formal communication, as well as the publishing of data,
its focus is on modelling the publishing and indexing of traditional
peer-reviewed journal articles, and finding and retrieving them. New
developments enabled by the Internet, such as open access journals and
e-print repositories, are also included.

A comparison with a previous model developed by Garvey and Griffith
in 1972 \cite{garvey} gives an idea of how recent developments 
added steps and facilities, but also highlighted the inner structure
of processes that in traditional publishing appeared elementary.

A good analysis has been made by Smith~\cite{smith} who identifies the
\emph{main roles}: editorial, quality cointrol (content), quality
control (form), recognition of work done, marketing, disseminating;
and the \emph{hidden roles}: subject defining, community defining,
archiving.

More simply Van de Sompel et al.~\cite{sompel-rethinking}, based
on~\cite{roosendaal}, distinguish the following five functions of
scholarly communication: Registration, Certification, Awareness,
Archiving and Rewarding.

It is also interesting to read Elsevier's point of view~\cite{dewaard}
focused on technologies (semantic Web) and publishers' projects.

\paragraph{The Deconstructed Journal.} 
The first important proposal was that of J. Smith \cite{smith,
smith-2004} of a Deconstructed Journal. DJ is based on a net of
Subject Focal Points (SFP), subject dedicated portals that link to
relevant items in the covered subjects. SFP's are different from
journals in that they don't own nor host the articles, but point to
them offering organized access and search facilities to
subscibers. SFP's do not organize quality control.

The latter role is played by ``evaluator organizations'', that are
paid by the author interested in having his work assessed. Nothing
prevents authors having thir work evaluated by more than one
evaluator, and multiple SFP's can point to the same item.

Archiving would be ensured either by local repositories or by not for
profit organisations like JSTOR~\cite{jstor}.

The author view would be: (1) Prepare the
article. (2) Place it on a visible server. (3) Notify one or more
evaluator organizations. (4) Revises it in the light of comments. (5)
Notify the relevant SFPs who place it on their recommended list if
it is relevant.

Smith claims that this model would solve some well known problems of
the traditional model such as the scattering of information across
many journals in the same field, easier publication for unconventional
ideas, delays in refereeing.

\paragraph{Two tier journal.} 
Another model has been proposed by Paul Ginsparg~\cite{ginsparg2002,ginsparg2004}.

First a minimal, maybe semi-automatic, filter to access the standard
tier. Then a much smaller set of articles is selected for the full
peer-review. The selection can be based on objective data (citation
impact, usage statistics) or user driven (reader nomination or rating) or
editorial. Review, being on a smaller number of articles, could be
combined: traditional, open or closed discussion \ldots\  Standard tier
could be made up of institutional repositories and/or disciplinary archives.

\paragraph{``The system that scholars deserve''.}

In the stream of the analysis conducted by Van de Sompel and
colleagues~\cite{sompel-rethinking,warner},
M. Rodriguez~\cite{rodriguez} proposed a deconstructed publication
model in which the peer review process is mediated by an
OAI\footnote{Open Archive Initiative, see
page~\pageref{oai}.}-compliant peer-review service. This peer-review
service uses a social-network algorithm to automatically determine
potential reviewers for a submitted manuscript and for weighting the
influence of each participating reviewer's evaluations.

The social-network is based on coauthorship links, and selection of
referees for a given article is made by automatic analysis of the
bibliography. All potential reviewers can submit their comments, and
give a score. Average scores are computed with adjusments.
This system could allow also evaluation of the reviewers.

What is very interesting in this system is the selection of reviewers:
a similar algorithm could be adopted also to assist editors in the
choice.

\bigskip

The weak point that is common to all theoretical proposals is that
they imagine a change of the system as a whole.  Moreover if
different models are possible then the best would be to see them
coexist, allowing everybody to choose.

\subsection{Concrete proposals}

\paragraph{Atmospheric Chemistry and Physics.}

Maybe the most interesting proposal is the one by\label{atm} 
U.~P\"oschl~\cite{poschl} (see also Koop and P\"oschl in
the Nature debate \cite{nature-debate-pr}) who devises a two-stage (or multi-stage)
publication processes with interactive peer review and public
discussion. Its applicability is demonstrated by the open access
journal Atmospheric Chemistry and Physics
(ACP)~\cite{atmos-chem-phys}. ACP is a lively journal: many
submissions, high impact factor.  It is open access at a modest
author-pays fee.\footnote{Very interestingly, not only authors of
published papers pay, but all those that are admitted to
discussion. This reduces consistently the pay-per-publish fees.}

More in detail. After basic filter by editor, articles are posted on a
child journal called Atmospheric Chemistry and Physics Discussion,
where it is accessible for discussion, during which the referees'
comments (anonymous or attributed), additional short comments by other
members of the scientific community (attributed) and the authors'
replies are also published in ACPD. Afterwards, the authors are given
a chance to revise their paper, which is then sent out for expert
review, in which the identity of referees is confidential. At this
stage referees are only supposed to return a verdict of `pass' or
`fail', as any more detailed feedback should already have been
received during the review period.  If accepted, the final revised
papers are published in ACP.

ACPD has its own ISSN, articles can be left there also if rejected and
can be cited.

\paragraph{Different levels journals.}  The B.E. Journals in Theoretical 
Economics~\cite{bejte} are three connected economics journal, with unified submission.

Each submission will be considered simultaneously by these three
quality-rated journals. A submission will, however, be accepted by at
most one of these journals. The editor, after standard peer review,
decides in which of the three journals it would be most appropriate.
\begin{itemize}
\item Advances in Theoretical Economics publishes articles that make
  significant advances;

\item  Contributions to Theoretical Economics publishes articles that make
  important contributions to specific literatures;

\item  Topics in Theoretical Economics publishes articles of interest to
  those working on specific topics;
\end{itemize}

A similar system is for The B.E. Journals in Economic Analysis \&
Policy (4 journals in one) and The B.E. Journals in Macroeconomics (4
journals in one).

\paragraph{Behavioral and {B}rain {S}ciences.}

A very ``old'' journal~\cite{bbs} that already existed
before the web (born in 1979).  Very relevant articles, after peer review, are
selected for open commentary. The editor invites a number of reviewers
to send their comments (at least 10). Comments are reviewed (sometimes
edited). Sent to the author who can reply (but not modify the
article). All this material is published together in the journal (a
unique pdf file). It has a very high impact factor.

{\bf Psycoloquy}~\cite{psycoloquy} was similar. The
always quoted~\cite{harnad-implementing} journal created by Stevan
Harnad used to publish comments together with articles, but hasn't
published anything since 2002.

\paragraph{Living Reviews.} 
As a different journal concept we can consider Living
Reviews \cite{living-rev}, three journals publishing only Review
articles, solicited from experts in the field by an international
Editorial Board, subject to peer-review, regularly being updated by
their authors to incorporate the latest developments in the field.
Nothing innovative in peer review, however. A similar model could be
used for a particular section of a journal, devoted to reviews.

\paragraph{PLoS ONE.} 
The new journal of the PLoS family~\cite{plosone}. The novelty is not
only in the access model, common to all PLoS journals, but in that it
is a unique big journal, no restricted to a subject, with open and
closed peer review.  It has just opened for submission (august~2006).

Each submission will be assessed by a member of the editorial
board before publication. This pre-publication peer review will
concentrate on technical rather than subjective concerns and may
involve discussion with other members of the editorial board and/or
the solicitation of formal reports from independent referees. If
published, papers will be accompanied by comments from the handling
editorial board member and will be made available for community-based
open peer review involving online annotation, discussion, and rating.
Reviewers may remain anonymous, but are
strongly urged to sign their reports.

\paragraph{InterJournal.}
Another journal based on the idea that more interdisciplinary
publishing areas are needed~\cite{interjournal}. Aims to cover various
topics in science and engineering. Moreover it does not directly host
full text articles --- that are left on the author choosen archive or
server, together with all accompanying material such as computer
programs, raw data, videos --- but only metadata and comments.

Manuscripts are immediately accessible upon submission, and any
qualified referee can review the article. However, at the discretion
of the authors, during an initial refereeing period access may be
limited to a few editor selected referees. The acceptance, category of
publication, and subject areas of publication are ultimately
determined by the editors.

Also {\bf Advances in {T}heoretical and {M}athematical
{P}hysics}~\cite{atmp} used to be an overlay of the archives and
provided only certification. After an interruption of the service it
is now a classic on-line journal.

\paragraph{Philica.} \cite{philica}
 is an online academic
journal accepting publications on any subject. It provides a process of
academic peer review, transparent (reviews can be seen
publicly) and dynamic (because opinions can change over time, and
this is reflected in the review process). 

Only academics can register and hence publish papers and write
reviews.

Philica is like eBay for academics. When somebody reviews your
article, the impact of that review depends on the reviewer's own
reviews. This means that the opinion of somebody whose work is highly
regarded carries more weight than the opinion of somebody whose work
is rated poorly. A person's standing, and so their impact on
other people's ratings, changes constantly. Reviews are anonymous.

\paragraph{Naboj.} Also Naboj~\cite{naboj} is not a proper journal. 
It lets readers review online scientific articles. At present it
allows to review only articles from {\tt arXiv}~\cite{arxiv}.  The review system is
modeled on Amazon and users have an opportunity to evaluate the
reviews as well as the articles. Alpha version. Not many posts so far.

\section{Peer review}\label{peerreview}

\subsection{What's wrong with peer review?}

Many studies have made a thorough analysis of all the problems of peer
review (see for instance Rowland~\cite{rowland},
McKiernan~\cite{mckiernan}, Williamson~\cite{williamson},
P\"oschl~\cite{poschl}, Grivell~\cite{grivell}).  In 2002 a general
review of 19 different studies~\cite{jefferson2002} concluded that
there is ``little evidence for effectiveness of scientific peer
review''.

A summary list of problems could be:
\begin{itemize}
\item \emph{Cost.} 

\item \emph{Subjectivity.} 

\item \emph{Bias.} Discrimination, or situations where author and
referee are competitors in some sense, or belong to warring schools of
thought.  There is evidence\footnote{Evidence means
statistical studies that report a correlation between review outcomes and
different categories.} of discrimination by fame, institution, geography, gender.  Some
institutions, like COPE \cite{cope}, try to vigilate and assist
complaining authors.

\item \emph{Abuse.} By authors
\begin{itemize} 
\item 
 Too many articles out of one piece of research (so called
 salami-pu\-bli\-shing), or duplicate publication. 

\item Aiming too high: authors aiming too high enter a downward spiral
of peer-review and rejection until the paper reaches its
level. The same paper is reviewed many times (multiplication
of costs, loss of time). 
To address this concern see \cite{bejte}

\item Intellectual theft: omission or downgrading of junior staff by
senior authors. 
\end{itemize}
and by reviewers
\begin{itemize}
\item Plagiarism (stealing others yet unpublished work that has been
sent for review). 

\item Delaying publication of potentially competing research.
\end{itemize}

\item \emph{Detecting defects.} Godlee et al.~\cite{jama-2} report an
experiment in which they modified a paper (already known and accepted
for publication) introducing 8 areas of weakness and sent it to 420
reviewers. In the 221 reports received an average of 2 defects was
detected. No one reported more than five defects, several zero.

\item \emph{Fraud and Misconduct.} Not generated by peer review but
almost impossible to detect for referees
\begin{itemize}
\item Fabrication of results 
\item Falsification of data 
\item False claim of authorship for results 
\end{itemize}

\item \emph{Delay.}
	Peer review is too slow, even though several journals set targets and
often obtain they are respected. 
\end{itemize}

Jennings in the Nature debate~\cite{nature-debate-pr} defends peer-review as to find
something better should meet several criteria.  Something could help
editors: manuscript tracking systems can provide feedback on where
delays arise; feedback on the quality of their decisions: look
retrospectively at the citations to accepted versus rejected papers,

In the end more than one author applied to peer review the famous
Churchill's sentence on democracy: it is the worst form of government
except all those other forms that have been tried from time to time.

\subsection{Alternative peer review models}

\paragraph{Neo-classical.} Classical peer review transplanted in the
World Wide Web. This is what the great majority of journals do,
particularly those that existed also as paper journals, with different
degrees of web integration. From almost email only, to all-in-one web
applications.

\paragraph{No peer review.}
There is a lot of people suggesting that peer review is no longer
necessary, starting from the observation that all the articles on {\tt
arXiv}~\cite{arxiv} are eventually published on some journal and that citation
figures of {\tt arXiv} itself and refereed journals are not
different~\cite{fabbrichesi-peer}.

S.~Mizzaro~\cite{mizzaro} proposes a system where all articles in a
given repository are voluntarily and freely rated by readers. A very
refined algorithm takes into account each reader's judgment weighted on
the basis of the reader's skill as a reviewer, and readers are
encouraged to express correct judgments by a feedback mechanism that
estimates their own quality. This system was designed for the Tips
project, see a simpler explanation in~\cite{tips-quality-2}.

S.~Harnad correctly points out that as long as peer review exists also
the free material is written with that in mind: the invisible hand of
peer review \cite{harnad-invisible}. Also R.~Kling~\cite{kling2004}
thoroughly analyzes the model of unrefereed scholarly publishing
highlighting problems and difficulties.

An example of journal without peer review could be the naive
experiment of Electronic Journal of Cognitive and Brain
Sciences~\cite{ejcbs} proposed by Z.~Nadasdy \cite{nadasdy}: articles
are freely rated by readers.

\paragraph{Open peer review.} The most concrete and experimented
proposal, although not a revolution. 

The \emph{British Medical Journal} \cite{BMJ} is probably the most
prestigious journal to make reviewers' names known to the
authors~\cite{bmj-open}.

In \emph{BioMed central}~\cite{BMC}, Many journals operate traditional anonymous
peer review. Others, including the medical BMC-series titles,
operate `open peer review', in which reviewers are asked to
sign their reviews. For these titles, the pre-publication
history of each paper (including submitted versions,
reviewers' reports and authors' responses) is linked to from
the published article.

In \emph{Journal of Medical Internet Research} \cite{jmir} names of
reviewers are published at the bottom of the paper.

The most complete analysis of the advantages of open peer review was
done by F.~Godlee \cite{godlee}: ethical superiority, accoutability,
 credit for the reviewers and authors' favour greatly outbalance
the adverse effects. As revealed by a thorough study published on JAMA
revealing the reviewers' identity doesn't seem to affect significantly
reviewers' recommendations or time taken to review~\cite{jama-1}
detection of errors~\cite{jama-2} and quality of reviews~\cite{jama-3,jama-4}.

\emph{Biology Direct}~\cite{bio-direct} (see also Koonin et al.\ in the
Nature debate~\cite{nature-debate-pr}) not only makes peer review open
but removes the journal's role in reviewer selection, making the
author responsible for obtaining three reviewers' reports, via the
journal's Editorial Board.  In essence, an article is rejected from
the journal if no appropriate Board member agrees to review it,
because agreeing means associating one's name to that article.
Limitations are imposed to avoid frequent author-referee couplings.

Linked to open peer review is the idea of authors suggesting one or
more reviewers' names (for instance in JMIR~\cite{jmir} and many other
journals) Also in this case quality of reviews is not significantly
affected according to E.~Wager et al.~\cite{wager-reviewers}.  Another
study points out that suggesting or excluding reviewers raises the
probability of publication~\cite{grimm}.

Many journals who maintain referees' names secret, at least acknowledge
them publishing every year the full list of referees.

\paragraph{Double-blind peer review.} In order to reinstate
equilibrium in the author-reviewer dynamic some suggest to hide the
author's identity as well~\cite{mainguy}. It is suggested that this
should address many concerns about peer review (subjectivity, bias
\dots\ see above). But this would work only with extremely honest
reviewers, otherwise it is very easy to retrieve author's name (in
physics it's enough to search the {\tt arXiv}es~\cite{arxiv}).

This kind of peer review is used, for instance, by \emph{{ASTRA} ---
{A}strophysics and {S}pace {S}ciences {T}ransactions}~\cite{astra}.

\paragraph{Commentary-based.}
\begin{itemize}
\item \emph{Commentaries pre PR:} this is the case already described
in \emph{Atmospheric Chemistry and Physics} \cite{atmos-chem-phys}. It was
also adopted by \emph{Electronic Transactions on Artificial Intelligence}
\cite{ETAI}: it provided a process for open discussion about articles
and feedback to authors before an article was accepted. This
discussion was shown and preserved on the ETAI website, and
participants in the discussion were not anonymous.  The discussion
about the article was combined with subsequent confidential refereeing
where referees are only supposed to return a verdict of `pass' or
`fail'.  The journal hasn't published since 2002

The noteworthiest experiment of this kind is the one by
\emph{Nature}~\cite{editorial-trial} that in june 2006 launched a trial of
open discussion parallel to traditional peer-review. A submitted
article, if the author so whishes, is posted in an open repository
where it can be freely discussed. Traditional peer review goes on in
parallel, and the editor can make the most of the discussion to
achieve his decision. Very few comments posted so far.

\item \emph{Commentaries post PR:} 
\emph{Psycoloquy} and \emph{Behavioural and Brain Sciences} have already
been mentioned in the journal model
section \cite{bbs,psycoloquy}.

The \emph{Medical Journal of Australia}~\cite{mja} in 1997 made a
trial for post acceptance commentary to be published along
with articles\\
{\small \href{http://www.mja.com.au/public/information/project.html}
{\tt
http://www.mja.com.au/public/information/project.html}}. No
report on results and no such issue in recent publications.

In \emph{Expert {R}eviews in {M}olecular {M}edicine}~\cite{expertrev}, each
article has an accompanying discussion group. Anyone can post comments
and questions and/or reply to previous comments, with the possibility
of being alerted for direct replies or new posts.

\item \emph{Commentaries pre \& post PR:} 
the \emph{Journal of Interactive Media in Education}~\cite{jime} has a
very complicate and thorough review process divided into stages:
submission, editor chooses three experts, experts discuss with author,
paper is made available to public and discussed by readers, authors
and experts, paper is revised, publised. Discussion can continue. All
the material is published alongside the paper.

Established in 1996, after a good initial success now publishes very
few articles.

\end{itemize}

\paragraph{Institution-based.} One example is the so called guild
model \cite{kling-guild}  based on the assumption that a
given department (or research unit) has its own manuscript
series. Access to these series is reserved to faculty members,
with no other filter. In some sense papers are judged based on
the careers of their authors.

Several softwares, like D-Space~\cite{dspace} and
eprints~\cite{eprints}, allow institutions to set up their manuscript
repositories.

A different situation is the one of large HEP collaborations, studied
by \cite{draper}, where are internal peer review is in place. When submitted
to peer reviewed journals, 100\% of papers are eventually accepted,

\paragraph{Citation-based.} Not really proposed or applied. Citation
analysis could assist editors in the choice of referees (see above the
journal that scholars deserve~\cite{rodriguez}).  A very interesting
tool that analyzes citations is CiteBase~\cite{citebase}.

Of course also other services like Scopus~\cite{scopus}, the (in)famous
Science Citation Index~\cite{sci} or Scholar
Google~\cite{scholargoogle} cover citation counting with different
outcomes and different usefulness~\cite{bakkalbasi}.

\paragraph{Computer-assisted.}
The Journal of High Energy Physic \emph{JHEP}~\cite{jhep} falls in
this category because of the automatic assignemt to editors (based on
a keywords driven algorithm). Such a system is not claimed by any other
journal, it would probably deserve to be more widely publicized.

This is probably a field where many advances can be envisaged.

The automatized assignment to the editor could be improved using
cognitive filtering: automatic analysis of a text (maybe the abstract
or the full text) can create a profile to be compared with similar
profiles created for each editor. This would give automatic
assignments without even havign to choose keywords (a potential source
of mistakes).

Editors could be helped in the choice of referees by social-network
based algorithm like the one proposed by Rodriguez~\cite{rodriguez}.

It would also be possible to provide editors and referees with
``objective data'' concerning the paper like download and citation figures
of the paper as a preprint.

\paragraph{Moderator-based.}
D.~Stern~\cite{stern} suggests a system where the moderator filters
out only inappropriate or offending papers. All others are
archived. Some (by the moderator, by download figures, by readers'
suggestion) are pointed to the attention of the editorial board.

In some sense it is similar to the model proposed by Rodriguez (see
above the ``system that scholars deserve'').

\section{News from the outside world}

W.~Arms~\cite{arms-quality} (see also Arms in the Nature
debate~\cite{nature-debate-pr}) invites to consider what examples the
web offers about quality assessing; even those coming from very
different contexts. Along this path we can mention:

\paragraph{Wikipedia.} On the footsteps of Wikipedia, an interactive 
encyclopaedia of the subject covered by the journal(s) could be
established.  Maybe a first version of some articles could be written
by invited scientists. There are many open source softwares to manage a
system like wikipedia (see for instance ``wiki''on Wikipedia).

Also a list of keywords, or the taxonomy to categorize the materials
could be managed collaboratively.
See Anderson in the Nature debate \cite{nature-debate-pr}.

\paragraph{Del.icio.us.} It is a web site~\cite{delicious} for the 
public archiving of bookmarks. A collaborative tagging systems
(analyzed by A.~Golder et al.~\cite{tagging-system}) entirely user
driven allows classification of the material.

Many other sites use collaborative tagging: \emph{Flickr} (images)
\emph{Dailymotion} (video), \ldots and also \emph{Connotea}~\cite{connotea}, the
Nature Publishing Group free online reference management system,
similar to Del.icio.us, for all researchers, clinicians and scientists
to publicly archive and classify scholarly references and bookmarks.

Particularly devoted to articles is \emph{CiteULike}~\cite{citeulike}. Always
with collaborative tagging. When one sees an interesting paper on the
web, he can click one button and have it added to the personal
library. CiteULike automatically extracts the citation details.  The
library can be shared with others, and one can find out who is reading
the same papers. In turn, this can help to discover relevant
literature.

\paragraph{Blogs.} {\sloppy Blogs are now quite frequent also in scientific 
communication. Nature has recently 
added blogs to its news articles (see for instance the blog on peer review
{\small\href{http://blogs.nature.com/nature/peerreview/debate/comments/}
{\tt http://blogs.nature.com/nature/peerreview/debate/comments/}}).

}

Concernig blogs see also~\cite{reichardt}. As an example of blog for
scientists see the \emph{Mass {S}pectrometry {B}log}~\cite{msb}.

The portal \emph{Postgenomic.com}~\cite{postgenomic} collates posts from life
science blogs and then does useful and interesting things with those
data. For example, you can see which papers are currently being
discussed by neurologists, or which web pages are being linked to by
bioinformaticians. It also uses a tagging system.

\paragraph{Internet drafts.}
As Arms~\cite{arms-quality} points out there are situations where the final
document must be almost perfect, and hence, without worrying about time,
very complicated review processes are set up.

As described in at the The Internet Engineering Task Force~\cite{ietf}
anybody can submit an Internet Draft, which will be 1. published as an
Internet Draft; 2. openly commented on; 3. revised by its author(s);
Steps 1 to 3 can be repeated a number of times.  Then an Area Director
has to take the draft to the IESG (Internet Engineering Steering
Group) for wider discussion. Further changes can be required, or the
draft can be rejected. The document still has to go through the states
of Proposed Standard and Draft Standard to eventually become a Full
Internet Standard. The whole procedure usually takes years.

A similar procedure is in place for the World Wide Web Consortium
Process Document.

The interest of these processes is in the fact that very high quality
is guaranteed without peer review. It could be of inspiration for the
production of particular documents, like collaborative review
articles.

\section{Standards to be compliant with}

These are the standards that nowadays all the most important journal respect.
\begin{itemize}
\item \label{oai} Open Archives Initiative~\cite{oai}. Not only open
access archives, but also big sub\-scrip\-tion-based publishers make their
metadata available in a OAI-PMH-compliant format. 

\item DOI~\cite{doi} --- Digital Object Identifiers are used to
retrieve objects persistently.  A resolver is like
CrossRef~\cite{crossref} is needed to transform DOIs into URLs:
CrossRef participating publishers deposit article metadata, and
CrossRef provides the link from the DOI to the original object.

A serious problem with DOI is explained in \cite{hurd}: when a library
has access to a resouce through a consortium or another web site, this
is ignored by DOI resolvers and the reader ends up in a page where
cannot read the full text.

All major publishers take part in DOI, including IOPP.

\item Unlike DOI, OpenURL~\cite{openurl} enables context-sensitive
linking from a reference in a scholarly information system to
resources relevant to the referenced item. Clicking on a link will
pass the request to a linking server enabling the provision of a list
of relevant services appropriate for the user. 

See also A.~Apps~\cite{apps}: why OpenURL?

\item Feed RSS. A system to keep users updated on what's going on on a
site. IOPP already uses it
{\small\href{http://www.iop.org/EJ/help/-topic=rss/journal/1126-6708}
{\tt http://www.iop.org/EJ/help/-topic=rss/journal/1126-6708}}.

\end{itemize}


\newcommand{\noopsort}[1]{} \newcommand{\printfirst}[2]{#1}
  \newcommand{\singleletter}[1]{#1} \newcommand{\switchargs}[2]{#2#1}
\raggedright

\end{document}